\documentclass[floatfix,showpacs,aps,a4paper]{revtex4}

 \usepackage{bm}
 \usepackage{amssymb}
 \usepackage{latexsym}
 \usepackage{amsfonts}
 \usepackage{epsfig}
 \usepackage{subfigure}
 \usepackage{color}
 \definecolor{darkblue}{rgb}{0,0,.5}
 \usepackage[linktocpage, colorlinks=true ,linkcolor=darkblue, citecolor=darkblue]{hyperref}
 \usepackage[all]{hypcap}

 \newcommand{\expval}[1]{\left< #1 \right>}

 \newcommand{\nn}{\nonumber\\}
 
 \newcommand{\f}[1]{\mbox{\boldmath$#1$}}

 \newcommand{\bea}{\begin{eqnarray}}
 \newcommand{\ea}{\end{eqnarray}}
 \newcommand{\eea}{\end{eqnarray}}

 \newcommand{\abs}[1]{{\left| #1 \right|}}

 \newcommand{\HB}{H_{\rm B}}
 \newcommand{\HI}{H_{\rm SB}}
 
 \newcommand{\RB}{\bar{\rho}_{\rm B}}
 
 \newcommand{\NB}{n_{\rm B}}
 \newcommand{\ii}{{\rm i}}
 \newcommand{\ee}{{\rm e}}

\begin{document}

\title{Single electron transistor strongly coupled to vibrations: Counting Statistics and Fluctuation Theorem}
\author{Gernot Schaller$^1$}\email{gernot.schaller@tu-berlin.de}
\author{Thilo Krause$^1$}
\author{Tobias Brandes$^1$}
\author{Massimiliano Esposito$^2$}
\address{$^1$ Institut f\"ur Theoretische Physik, Technische Universit\"at Berlin, Hardenbergstr. 36, D-10623 Berlin, Germany}
\address{$^2$ Complex Systems and Statistical Mechanics, University of Luxembourg, L-1511 Luxembourg, Luxembourg}

\begin{abstract}
Using a simple quantum master equation approach, we calculate the Full Counting Statistics of a single electron transistor strongly 
coupled to vibrations. The Full Counting Statistics contains both the statistics of integrated particle and energy currents associated to the transferred electrons 
and phonons. A universal as well as an effective fluctuation theorem are derived for the general case where the various reservoir 
temperatures and chemical potentials are different. The first relates to the entropy production generated in the junction while the second 
reveals internal information of the system.
The model recovers Franck-Condon blockade and potential 
applications to non-invasive molecular spectroscopy are discussed.
\end{abstract}


\maketitle

The Full Counting Statistics (FCS) of energy and matter exchanges provides a wealth of information about the dynamics of multi-terminal nanostructures. 
A great achievement of the last decade has been to conveniently
identify universal features in the FCS~\cite{nazarov2005,esposito2009a,andrieux2009a,utsumi2009a,campisi2011a,utsumi2013a}.
Roughly speaking, these are related to the fact that fluctuations of entropy production, $\Delta_{\rm i} S$, satisfy a 
universal fluctuation theorem (FT) $P_{+\Delta_{\rm i} S}/P_{-\Delta_{\rm i} S} = e^{\Delta_{\rm i} S}$.
At steady state, the entropy production in the nanostructure must be balanced by the 
entropy flow through its terminals~\cite{lebowitz1999a}, which can be accessed through the FCS.
This implies that entropy production is a measurable quantity and can be expressed as a sum of the 
various thermodynamic affinities acting on the system times their associated fluxes~\cite{andrieux2007a}.
Fostered by the increased experimental abilities enabling the counting of single electron transfers~\cite{ExpFT1} and thereby the experimental verification of the FT~\cite{ExpFT2}, recent works have identified non-universal FT-symmetries when an incomplete 
monitoring of all the terminals is performed. 
The resulting effective affinities may be used to probe system-specific 
features~\cite{EffectiveFT1,EffectiveFT2,EffectiveFT3,EffectiveFT4,EffectiveFT5}.

Nanoscale devices displaying coupling between electronic and vibrational transport have been significantly studied in the past, 
in part due to their importance for thermo-electricity \cite{ThermoElecWC1,ThermoElecWC2,ThermoElecWC3}. 
Most studies rely on the weak coupling assumption and those who don't use sophisticated self-consistent 
nonequilibrium Greens functions procedures which often restrict the study of the FCS to the first few moments 
\cite{galperin1,galperin2,haupt,park}.

In this paper, we calculate the FCS for a single electron transistor (SET) strongly coupled to a phonon bath using a new particularly simple quantum master equation (QME) approach.
We explicitly derive the universal FT in the general case when different electronic and phononic temperatures are considered.
Effective FTs are also identified and both the electron and phonon average currents and Fano factors are analysed.
Going beyond heat exchange~\cite{nicolin1,nicolin2}, our model provides an efficient probe for non-destructive molecular spectroscopy:
While it is known that one can determine phonon energies from electronic transport characteristics only, the simplicity
of our model allows to identify parameter regimes where this is possible without inducing further heating of phonons.


\section{Model}
We consider the Hamiltonian
\bea 
H &=& \epsilon d^\dagger d + \sum_{k\alpha} \epsilon_{k\alpha} c_{k\alpha}^\dagger c_{k\alpha} + \sum_q \omega_q a_q^\dagger a_q\nn
&&+ \sum_{k\alpha} \left[t_{k\alpha} d c_{k\alpha}^\dagger + {\rm h.c.}\right]+ d^\dagger d \sum_q \left[h_q a_q + {\rm h.c.}\right]
\eea
describing a SET with two electronic leads (treated perturbatively) and additionally coupled to one or many phononic modes (treated non-perturbatively). 
Fermionic operators $d$ ($c_{k\alpha}$) annihilate electrons on the dot (lead $\alpha\in\{L,R\}$) with energies $\epsilon$ ($\epsilon_{k\alpha}$), and 
$a_q$ are the bosonic annihilation operators for a phonon with energy $\omega_q$.
The parameters $t_{k\alpha}$ and $h_q$ describe electronic tunneling and phononic absorption amplitudes, respectively.
For only a few phonon modes $q\in\{1,\ldots,Q\}$, the model may e.g. describe electronic transport through a molecule where the electronic occupation couples to molecular vibrations.
In particular for a single phonon mode ($Q=1$), the model represents a special case of the Anderson-Holstein model, 
which has previously been treated for example in the linear response and weak electron-phonon-coupling regimes~\cite{entin_wohlman2010a} or
with focus on the electronic charge transfer statistics~\cite{maier2011a,white2012a}.
For many different phonon modes or even a continuum, the model may describe more complex molecules or the interaction with bulk phonons, respectively.

Establishing the complete electronic and phononic FCS requires to monitor not only the charge transfer but also
the emitted and absorbed bosons.
In principle, the statistics of the latter can be retrieved by modelling the phonons as part of the 
system, see e.g. Refs.~\cite{koch2005a,segal2008b,avriller2011a,mitra2004a}.
This allows to explore the strong electron-phonon coupling limit, but also leads to an infinitely large Hilbert space, which renders the study of the full dynamics tedious and hard to interpret. 
For many different phonon modes or even a continuum the required computational resources make it completely infeasible to follow this approach.
Here, we therefore aim at an efficient representation with only the dot occupation treated as a dynamical variable whilst all 
electronic and phonon terminals are held at thermal equilibrium states.
The challenge is to retain the complete FCS from such a reduced model.
We would like to emphasize here that even in the single phonon mode case the postulated stationarity of the phonons does not imply a weak coupling assumption between electrons and phonons.
For example, for strongly coupled phonons there might exist an even faster relaxation process for the phonons immediately restoring thermal equilibrium.

After a polaron (Lang-Firsov) transformation~\cite{mahan1990,brandes2005} \mbox{$H'=e^{+S} H e^{-S}$} with
\mbox{$S = d^\dagger d \sum_q (\frac{h_q^*}{\omega_q} a_q^\dagger - \frac{h_q}{\omega_q} a_q)$}, where
\bea
H'&=&\tilde\epsilon d^\dagger d + \sum_{k\alpha} \epsilon_{k\alpha} c_{k\alpha}^\dagger c_{k\alpha} + \sum_q \omega_q a_q^\dagger a_q\nn
&&+\sum_{k\alpha} \left[t_{k\alpha} d c_{k\alpha}^\dagger e^{-\sum_q \left(\frac{h_q^*}{\omega_q} a_q^\dagger - \frac{h_q}{\omega_q} a_q\right)} + {\rm h.c.}\right]\,,
\eea
the dot energy is renormalized $\tilde\epsilon\equiv \epsilon-\sum_q \frac{\abs{h_q}^2}{\omega_q}$, 
and the electronic tunneling is dressed by exponentials of bosonic annihilation and creation operators.
Expanding these terms demonstrates that each electronic tunneling event may now be accompanied by 
multiple phonon emissions and absorptions.
Most important however, we note that a perturbative treatment in the electronic tunneling amplitudes $t_{k\alpha}$ -- valid for small electronic tunneling rates in comparison
to the electronic reservoir temperatures -- still allows for
a non-perturbative treatment of the electron-phonon interaction (parametrised by $h_q$).

We now apply standard techniques (e.g. \cite{esposito2009a,schaller2009b}) to derive a QME containing the FCS of both 
the emitted/absorbed phonons and the electrons having traversed the system. 
Note that since we are perturbative in the electronic tunneling amplitudes, we neglect coherences and also the Kondo effect remains out of reach.
For a system-bath decomposition of the form $\HI = \sum_i A_i B_i$ with system and bath operators $A_i$ 
and $B_i$, respectively, this QME requires to calculate the bath correlation function
$C_{ij}(\tau) = \expval{e^{+\ii \HB\tau} B_i e^{-\ii\HB\tau} B_j}$, where the expectation value is taken 
with respect to the stationary reservoir state~\cite{breuer2002}.
In our case, the latter consists of a tensor product $\RB = \RB^{(L)} \otimes \RB^{(R)} \otimes \RB^{(\rm ph)}$ 
of different equilibrium states characterized by temperatures, and for the electronic leads also by chemical potentials.
We choose the system coupling operators as $A_{1,\alpha}=d$ and $A_{2,\alpha}=d^\dagger$ and correspondingly we see that 
the contribution from the phonon bath enters multiplicatively in the bath coupling operators
\mbox{$B_{1,\alpha} = \sum_{k} t_{k\alpha} c_{k\alpha}^\dagger \exp[-\sum\limits_q (\frac{h_q^*}{\omega_q} a_q^\dagger - \frac{h_q}{\omega_q} a_q)]$} and 
\mbox{$B_{2,\alpha}=B_{1,\alpha}^\dagger$}.
In contrast, the contributions from the two electronic leads enter additively as usual.
This leads (see~\ref{SAbcf} for more details) to a non-standard product form of the correlation function
$C_{\ell}(\tau) = \sum_\alpha C_{\ell}^{\alpha,\rm el}(\tau) C_{\rm ph}(\tau)$, where $\ell \in \{(12),(21)\}$.
For the electronic contribution to the correlation function, we have the usual Fourier decomposition 
\mbox{$C_{\ell}^{\alpha, \rm el}(\tau) = \frac{1}{2\pi} \int \gamma_{\ell}^{\alpha,\rm el}(\omega) e^{-\ii\omega\tau} d\omega$}
with the standard electronic Fourier transforms
\bea\label{Ebcfset}
\gamma_{12}^{\alpha,\rm el}(\omega) = \Gamma_\alpha(-\omega) f_\alpha(-\omega)\,,\qquad
\gamma_{21}^{\alpha,\rm el}(\omega) = \Gamma_\alpha(+\omega)\left[1-f_\alpha(+\omega)\right]\,,
\eea
where $f_\alpha(\omega)\equiv\left[e^{\beta_\alpha(\omega-\mu_\alpha)}+1\right]^{-1}$ denotes the Fermi function of lead $\alpha$ held
at inverse temperature $\beta_\alpha$ and chemical potential $\mu_\alpha$. 
The electronic tunneling rates in Eq.~(\ref{Ebcfset}) are defined by \mbox{$\Gamma_\alpha(\omega) = 2\pi \sum_k \abs{t_{k\alpha}}^2 \delta(\omega-\epsilon_{k\alpha})$}.
In the following, we will parameterize them by a Lorentzian shape $\Gamma_\alpha(\omega) = \Gamma_\alpha \delta^2/(\omega^2+\delta^2)$ with
$\delta\to\infty$ characterizing the wide-band limit.
Finite $\delta$ may effectively model non-Markovian effects induced by the electronic environment~\cite{zedler2009a}.
Combining the phonon contributions to the bath correlation function using the Baker-Campbell-Hausdorff formula, we obtain
\bea\label{Ebcfphonon}
C_{\rm ph}(\tau) = 
\exp\left\{\sum\limits_q \frac{\abs{h_q}^2}{\omega_q^2}\left[e^{-\ii\omega_q\tau}\left(1+\NB^q\right)+e^{+\ii\omega_q\tau} \NB^q-\left(1+2\NB^q\right)\right]\right\}
\eea
with $\NB^q \equiv \left[e^{\beta_{\rm ph} \omega_q}-1\right]^{-1}$ denoting the Bose distribution with inverse temperature $\beta_{\rm ph}$.
We note that the Fourier transform of the combined correlation function 
\mbox{$\gamma_\ell(\omega) \equiv \int C_\ell(\tau) e^{+\ii\omega\tau} d\tau$} 
constitutes the transition rates in the rate equation when evaluated at the renormalized dot energies $\pm\tilde{\epsilon}$.


\section{Fluctuation Theorems}

\subsection{Entropy Production}

For a finite number of phonon modes, the FT for entropy production may be expressed in terms of the FCS, which is detailed in~\ref{SAfcsft}.
This is most accessible for a single phonon mode ($Q=1$) at frequency $\omega_1=\Omega$ 
(we also abbreviate $\NB^1 \to \NB$ and $h_1\to h$), where the sum in the exponential of Eq.~(\ref{Ebcfphonon}) collapses.
Then, one can identify the Fourier transform $\gamma_\ell^\alpha(\omega)$ of the correlation function via a simple integral transformation.
In particular, it is straightforward to see (see~\ref{SAbcf}) that it can be decomposed $\gamma_{\ell}^\alpha(\omega) = \sum_{n=-\infty}^{+\infty} \gamma_{\ell,n}^\alpha(\omega)$ into processes associated with the net absorption (emission) of $n>0$ ($n<0$) quanta by the phonon modes
\bea\label{Ecfunc}
\gamma_{\ell,n}^\alpha(\omega) &=& \gamma_{\ell}^{\alpha,\rm el}(\omega-n\Omega) e^{-\frac{\abs{h}^2}{\Omega^2}\left(1+2\NB\right)}\left(\frac{1+\NB}{\NB}\right)^{n/2}\times\nn
&&\times {\cal J}_n \left(2\frac{\abs{h}^2}{\Omega^2} \sqrt{\NB(1+\NB)}\right)\,,
\eea
where ${\cal J}_n(x)$ denotes the modified Bessel function of the first kind.
First, in the zero-coupling limit $\abs{h}^2\to 0$, we recover the correlation functions of the SET~(\ref{Ebcfset}) as all contributions
with $n \neq 0$ vanish.
Second, in the zero-phonon-temperature limit $\NB \to 0$, only contributions for absorption by the phonon bath remain with
\mbox{$\lim\limits_{\NB\to0} [\frac{1+\NB}{\NB}]^{n/2} {\cal J}_n(2 \frac{\abs{h}^2}{\Omega^2}\sqrt{\NB(1+\NB)})\stackrel{n\ge 0}{=}(\frac{\abs{h}^2}{\Omega^2})^n \frac{1}{n!}$}.
Third, in the wide-band ($\delta\to\infty$ such that $\Gamma_\alpha(\omega) \to \Gamma_\alpha$) and infinite bias limit ($\mu_{\alpha} \to \{-\infty,\infty\}$ so that $f_\alpha(\omega) \to \{0,1\}$), we recover the standard infinite-bias results of the SET.

\begin{figure}[ht]
\begin{center}
\includegraphics[width=0.75\textwidth,clip=true]{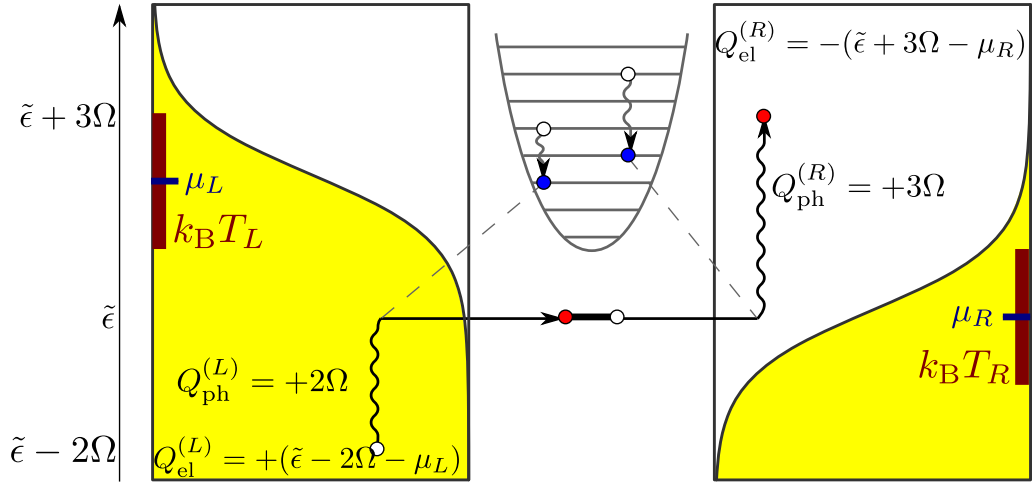}
\end{center}
\caption{\label{Fsketch_setvib}
Sketch of the heat transfer energetics at the trajectory level for the single-phonon-mode ($Q=1$) case and 
for two sample elementary processes (black solid with specific heat flows into the system noted).
Filled regions left and right denote Fermi functions of left and right electronic leads whereas the sketched oscillator denotes the
phonon reservoir, respectively.
During the electronic tunneling events (wavy lines), energy from the phonons may assist
electrons far from the dot level $\tilde\epsilon$ to participate in transport.
In the left trajectory, an electron jumps from the left lead to the initially empty dot via the emission of two quanta by the phonons.
In the right trajectory, it leaves the dot to the right lead via the emission of two quanta by the phonons.
For trajectories with identical initial and final SET state (the change of SET entropy is zero), the total entropy production is minus the entropy flow and is given by $\Delta_{\rm i} S = -Q_{\rm el}^{(L)}/T_L - Q_{\rm el}^{(R)}/T_R - (Q_{\rm ph}^{(L)}+Q_{\rm ph}^{(R)})/T_{\rm ph}$, with transferred electronic heat $Q_{\rm el}^{(\alpha)} = - (\tilde\epsilon n_{\rm el}^{(\alpha)} - e_{\rm ph}^{(\alpha)})+\mu_\alpha n_{\rm el}^{(\alpha)}$ and associated phonon heat $Q_{\rm ph}^{(\alpha)} = - e_{\rm ph}^{(\alpha)}$.
For example, the combination of both trajectories shown yields with $n_{\rm el}^{(L)}=-1$, $n_{\rm el}^{(R)}=+1$, $e_{\rm ph}^{(L)}=-2\Omega$, and $e_{\rm ph}^{(R)}=-3\Omega$ altogether a total entropy production of $\Delta_{\rm i} S = -(\tilde{\epsilon}-2\Omega-\mu_L)/T_L-5\Omega/T_{\rm ph}+(\tilde{\epsilon}+3\Omega-\mu_R)/T_R$.
}
\end{figure}

Both the Fermi-Dirac distribution and the Bose-Einstein distribution obey the separate relations
\mbox{$f_\alpha(\omega) = e^{-\beta_\alpha(\omega-\mu_\alpha)} \left[1-f_\alpha(\omega)\right]$} and 
$\NB = e^{-\beta_{\rm ph}\Omega} \left(1+\NB\right)$.
These imply that the correlation functions~(\ref{Ecfunc}) obey a relation of Kubo-Martin-Schwinger (KMS) type involving both electronic and phononic temperatures, 
\bea\label{Ekms}
\gamma_{12,+n}^\alpha(-\omega) &=& e^{-\beta_\alpha(\omega-\mu_\alpha+n\Omega)} e^{\beta_{\rm ph} n\Omega} \gamma_{21,-n}^\alpha(+\omega)\,,
\eea
which in the case of equal temperatures $\beta_{\rm ph}=\beta_L=\beta_R$ reduces to the conventional KMS condition~\cite{weiss1993}.
Evaluated at the renormalized dot level $\tilde{\epsilon} = \epsilon - \abs{h}^2/\Omega$, these enter the Liouvillian ${\cal L}$ 
of the QME describing the dot dynamics: $\dot{\rho} = {\cal L} \rho$, where the part acting on the populations only reads
\bea\label{Elmatfull}
{\cal L} &=&\sum_{\alpha\in\{L,R\}} \sum_{n_\alpha=-\infty}^{+\infty}
 \left(\begin{array}{cc}
-\gamma_{12,n_\alpha}^\alpha & \gamma_{21,n_\alpha}^\alpha e^{+\ii\chi_\alpha+\ii n_\alpha \Omega \xi_\alpha}\\
+\gamma_{12,n_\alpha}^\alpha e^{-\ii\chi_\alpha+\ii n_\alpha \Omega \xi_\alpha} & -\gamma_{21,n_\alpha}^\alpha
\end{array}\right)\,.
\eea
Here, $\gamma_{12,n_\alpha}^\alpha\equiv\gamma_{12,n_\alpha}^\alpha(-\tilde{\epsilon})$ denotes the transition rate from an empty to a filled dot due to an electron jumping in from lead $\alpha$ whilst simultaneously triggering the absorption of $n_{\alpha}$ quanta by the phonons.
Similarly, $\gamma_{21,-n_\alpha}^\alpha\equiv\gamma_{21,-n_\alpha}^\alpha(+\tilde{\epsilon})$ denotes the rate of the inverse process.
Effectively, this corresponds to transport through an electronic system with infinitely many reservoirs with shifted chemical potentials, and the energetics of heat transfers is easily accessible on the level of single trajectories, see Fig.~\ref{Fsketch_setvib}.
%
In order to monitor the full entropy flow through the junctions, it is in general necessary to monitor both energy and particle flows.
The electronic counting fields $\chi_{\alpha}$ count the net number of electrons $n_{\rm el}^{(\alpha)}$ jumping out of the SET to the lead $\alpha$, while the phonon counting fields $\xi_\alpha$ count the corresponding net energy transfered to the phonons $e_{\rm ph}^{(\alpha)}=n_\alpha \Omega$.
Together they can be used to determine the energy-particle FCS of our model, see~\ref{SAfcsft}, i.e., to calculate the time-dependent probability distribution 
\mbox{$P_{n_{\rm el}^{(R)},e_{\rm ph}^{(L)},e_{\rm ph}^{(R)}}(t)$} of electronic matter and associated phonon energy transfers.

Eq.~(\ref{Elmatfull}) is a key result of our paper which clearly shows that when phonons are kept thermally equilibrated, a compact description of the dynamics can be established which keeps the full information about the electron and phonon counting statistics also in the strong electron-phonon coupling regime.
In addition to thermalized phonons, the perturbative treatment of the electronic tunneling requires that $\beta_{\alpha} \Gamma_{\alpha} \ll 1$.
Our approach is consistent thermodynamically since as shown in~\ref{SAfcsft},
the following universal FT can be derived (also for multiple phonon modes)
\bea\label{Efttotal}
\ln \bigg( \frac{P_{+n_{\rm el}^{(R)},+e_{\rm ph}^{(L)},+e_{\rm ph}^{(R)}}(t)}{P_{-n_{\rm el}^{(R)},-e_{\rm ph}^{(L)},-e_{\rm ph}^{(R)}}(t)} \bigg) 
&\stackrel{t\to \infty}{=}& A n_{\rm el}^{(R)} + A_L e_{\rm ph}^{(L)} + A_R e_{\rm ph}^{(R)}\,,
\eea
where the affinities corresponding to the phonon energy and electronic fluxes are respectively given by
\bea
A_\alpha=\beta_{\rm ph}-\beta_\alpha\,,\qquad
A=(\beta_R-\beta_L) \tilde{\epsilon} + (\beta_L \mu_L - \beta_R \mu_R)\,.
\eea

We now turn to the discussion of Eq.~(\ref{Efttotal}). Each elementary transfer process described by QME~(\ref{Elmatfull}) involves an electron transfer between the SET and a lead $\alpha$ coupled to an energy transfer with the phonons. Since energy is conserved during such transfers, the energy that when $n_{\rm el}^{(\alpha)}>0$ leaves
the SET, $\tilde{\epsilon} n_{\rm el}^{(\alpha)}$, is equal to the energy sent to the phonons, $e_{\rm ph}^{(\alpha)}$, plus the energy sent to lead $\alpha$, $\tilde{\epsilon} n_{\rm el}^{(\alpha)}-e_{\rm ph}^{(\alpha)}$. As a result the heat entering the SET from the phonons is given by $Q_{\rm ph}^{(\alpha)} = - e_{\rm ph}^{(\alpha)}$ and the heat entering the SET from lead $\alpha$ by $Q_{\rm el}^{(\alpha)} = -(\tilde\epsilon n_{\rm el}^{(\alpha)} - e_{\rm ph}^{(\alpha)})+\mu_\alpha n_{\rm el}^{(\alpha)}$.
For a single phonon mode, $e_{\rm ph}^{(\alpha)}= n_\alpha \Omega$, while for multiple modes, $e_{\rm ph}^{(\alpha)}=\sum_q \omega_q n_{\alpha,q}$, where $n_\alpha$ and $n_{\alpha,q}$ denote the number of quanta with frequency $\Omega$ and $\omega_q$ that are emitted from the phonons (see~\ref{SAfcsft} for details).
The entropy flow associated to this process (corresponding to minus the change in the entropy of the phonon and that of the lead $\alpha$) will be given by $\Delta_\ee S^{(\alpha)} = \beta_{\alpha} Q_{\rm el}^{(\alpha)} + \beta_{\rm ph} Q_{\rm ph}^{(\alpha)}$~\cite{esposito2009a}.
At steady state, in average and in a large deviation sense (and in a strict sense for trajectories connecting identical initial and final SET states), entropy production is equal to minus the entropy flow, i.e. $\Delta_\ii S=- \sum_{\alpha} \big( \beta_{\alpha} Q_{\rm el}^{(\alpha)} + \beta_{\rm ph} Q_{\rm ph}^{\alpha}$ \big). 
Also, the total number of transferred electrons has to be conserved $n_{\rm el}^{(L)} + n_{\rm el}^{(R)}=0$.
As a result, we easily verify that the entropy production $\Delta_\ii S$ becomes equal to Eq.~(\ref{Efttotal}).

The FT~(\ref{Efttotal}) thus states that trajectories with a positive entropy production occur with larger probabilities than the inverse trajectories, i.e., the expectation value of the entropy production is always positive as predicted by the second law.
This holds far from equilibrium and non-perturbatively in the electron-phonon coupling strength.
Single trajectories -- occurring with an exponentially suppressed probability -- however may have a negative entropy production.
For example, when $\mu_L=\mu_R$, $\beta_L=\beta_R=\beta_{\rm el}$, and $\beta_{\rm ph} > \beta_{\rm el}$, the combination of the two elementary processes depicted in 
Fig.~\ref{Fsketch_setvib} with $n_{\rm el}^{(R)}=+1$, $e_{\rm ph}^{(L)}=-2\Omega$, and $e_{\rm ph}^{(R)}=-3\Omega$ would yield a trajectory with negative production.

As further consistency tests of Eq.~(\ref{Efttotal}) we mention that 
when $\beta_{\rm el} = \beta_R=\beta_L$, we get $A=\beta_{\rm el} (\mu_L-\mu_R)$ and $A_\alpha=\beta_{\rm ph}-\beta_{\rm el}$.
When furthermore $\beta \equiv \beta_{\rm el}=\beta_{\rm ph}$, $A_\alpha=0$ and the entropy production becomes identical to that of an isolated junction, i.e., $\Delta_{\rm i} S \to \beta V_{\rm bias} n_{\rm el}^{(R)}$ with bias voltage $V_{\rm bias} \equiv \mu_L-\mu_R$.
All affinities obviously vanish at equilibrium where $\beta_L=\beta_R=\beta_{\rm ph}$ and $\mu_L=\mu_R$.

Finally, we mention that when phonons were treated as part of the system, the derivation of the FT would be standard 
due to the additivity of the electronic leads:
It would not involve entropy flow across a strongly-coupled terminal and thus contain no phonon-related affinity.

\subsection{Incomplete Fluctuation Theorem}

\begin{figure}[ht]
\begin{center}
\includegraphics[width=0.75\textwidth,clip=true]{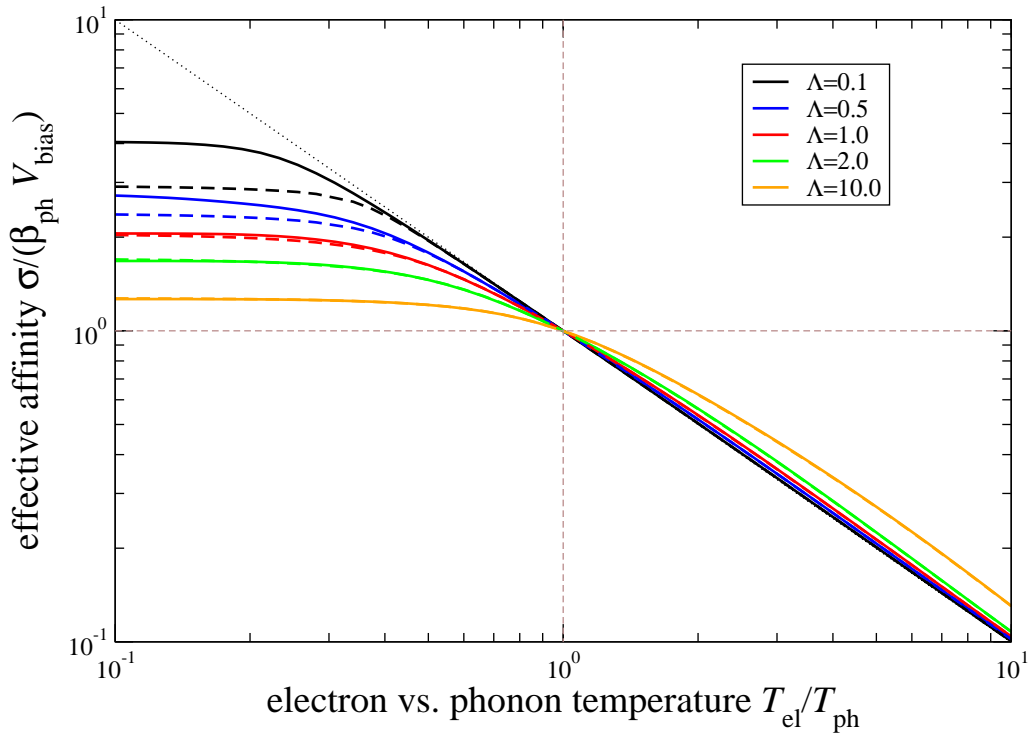}
\end{center}
\caption{\label{Fredelft}
Effective affinity $\sigma$, Eq.~(\ref{Esmftreduced}), for different values of the coupling strength $\Lambda=\abs{h}^2/\Omega^2$ in the single-mode case (solid lines) or $\Lambda=J_0$ in the continuum case (dashed curves).
When $T_{\rm el}=T_{\rm ph}$, the FT is independent of the coupling strength (intersection point).
For vanishing coupling strength, we recover $\sigma=\beta_{\rm el} (\mu_L - \mu_R)$ (dotted thin curve).
Other Parameters: $\Gamma_L = \Gamma_R$, $\beta_{\rm ph} \mu_L=+5$, $\beta_{\rm ph} \mu_R=-5$, $\delta=10\Omega$, $\beta_{\rm ph} \epsilon=0$, and $\beta_{\rm ph}\Omega=\beta_{\rm ph}\omega_c=1$.
}
\end{figure}

Returning to the general case but disregarding now the phonon heat counting, which is technically performed by setting $\xi_\alpha=0$, an incomplete FT~\cite{EffectiveFT1, EffectiveFT2, EffectiveFT3, EffectiveFT4, EffectiveFT5} is obtained for the probability $P_{n_{\rm el}^{(R)}}(t)$ of $n_{\rm el}^{(R)}$ electrons having crossed the junction after time $t$
\bea\label{Esmftreduced}
\ln\left(\frac{P_{+n_{\rm el}^{(R)}}(t)}{P_{-n_{\rm el}^{(R)}}(t)}\right) \stackrel{t \to \infty}{=} n_{\rm el}^{(R)} \cdot \sigma\,,\;\;\;
\sigma = \ln \left[ 
\frac{\gamma_{12}^L(-\tilde{\epsilon}) \gamma_{21}^R(+\tilde{\epsilon})}
{\gamma_{12}^R(-\tilde{\epsilon}) \gamma_{21}^L(+\tilde{\epsilon})}  \right]\,,
\eea
where the effective affinity $\sigma$ is not universal anymore (unless $\beta_L=\beta_R=\beta_{\rm ph}$) and can be evaluated numerically, see Fig.~\ref{Fredelft}.
Consequently, for an incomplete FT $n_{\rm el}^{(R)} \cdot \sigma$ cannot be associated anymore with the total entropy production.
However, the non-universal $\sigma$ does now provide additional information about the system: In the example shown in Fig.~\ref{Fredelft}, it may e.~g.~be used as an effective thermometer for the electron temperature $T_{\rm el}$ at fixed phonon temperature $T_{\rm ph}$.
Note that flat ($\delta \to \infty$) electronic tunneling rates cancel in Eq.~(\ref{Esmftreduced}).

For a quantum dot coupled to {\em bulk} phonons (continuum of modes), obtaining the resolved FCS of different modes is realistically impossible in an experiment.
Therefore, monitoring only the electronic statistics will also yield an incomplete FT.
Then, the sum in Eq.~(\ref{Ebcfphonon}) can be converted into an integral by introducing the 
spectral coupling density $J(\omega)=\sum_q \abs{h_q}^2 \delta(\omega-\omega_q)$.
The common choice of an Ohmic parameterization~\cite{brandes2005}
$J(\omega) = J_0 \omega e^{-\omega/\omega_c}$
with the dimensionless coupling strength $J_0$ and cutoff frequency $\omega_c$
enables one to calculate the phonon contribution to the correlation function~(\ref{Ebcfphonon}) analytically
(we use an overbar to denote the continuum case)
\bea\label{Ecfuncbosonic}
\bar{C}_{\rm ph}(\tau) = 
\left[\frac{\Gamma\left(\frac{1+\beta_{\rm ph} \omega_c + \ii \tau \omega_c}{\beta_{\rm ph} \omega_c}\right)
\Gamma\left(\frac{1+\beta_{\rm ph} \omega_c - \ii \tau \omega_c}{\beta_{\rm ph} \omega_c}\right)}
{\Gamma^2\left(\frac{1+\beta_{\rm ph} \omega_c}{\beta_{\rm ph} \omega_c}\right)(1+\ii\tau\omega_c)}\right]^{J_0}\hspace{-0.25cm},
\eea
where $\Gamma(x)$ denotes the $\Gamma$-function.
It is straightforward to verify a separate KMS condition in the time-domain
\mbox{$\bar{C}_{\rm ph}(\tau) = \bar{C}_{\rm ph}(-\tau-\ii\beta_{\rm ph})$}, 
which implies
$\bar{\gamma}_{\rm ph}(\omega) = e^{+\beta_{\rm ph} \omega} \bar{\gamma}_{\rm ph}(-\omega)$ in the frequency domain.
The full Fourier transforms of the correlation functions are therefore given by a convolution integral of the separate Fourier transforms
\mbox{$\bar{\gamma}_{\ell}^\alpha(\Omega)=\frac{1}{2\pi} \int \gamma_{\ell}^{\alpha, \rm el}(\omega) \bar{\gamma}_{\rm ph}(\Omega-\omega) d\omega$}.
Using the separate electronic and phonon KMS conditions demonstrates that for equal temperatures, the conventional KMS condition must also hold
in the continuum case.
In addition, in the wide-band and infinite bias limits -- where the electronic Fourier transforms~(\ref{Ebcfset}) become
constants (over a sufficiently wide range) -- the convolution representation implies that the corresponding electronic currents 
are unaffected by the additional phonon reservoir.
When only electrons are counted, these enter the QME as
\bea\label{Eredlmat}
{\cal L} &=& \sum_\alpha \left(\begin{array}{cc}
-\bar{\gamma}_{12}^\alpha(-\tilde{\epsilon}) & +\bar{\gamma}_{21}^\alpha(+\tilde{\epsilon})e^{+\ii\chi_\alpha}\\
+\bar{\gamma}_{12}^\alpha(-\tilde{\epsilon})e^{-\ii\chi_\alpha} & -\bar{\gamma}_{21}^\alpha(+\tilde{\epsilon})
\end{array}\right)\,.
\eea
Again it suffices to consider only a single electronic counting field to obtain an incomplete FT as 
in Eq.~(\ref{Esmftreduced}) with $\gamma_{\ell}^\alpha \to \bar{\gamma}_{\ell}^\alpha$, where the renormalized dot level
now becomes $\tilde{\epsilon} = \epsilon - J_0 \omega_c$.
When we choose different temperatures for electrons and phonons, 
the affinity $\sigma$ in the FT may be evaluated numerically, and displays a 
similar dependence as in the single-mode case provided $J_0=h^2/\Omega^2$ and $\omega_c=\Omega$,
cf.~dashed lines in Fig.~\ref{Fredelft} and the discussion below.


\section{Counting Statistics}

\begin{figure}[ht]
\begin{center}
\includegraphics[width=0.75\textwidth,clip=true]{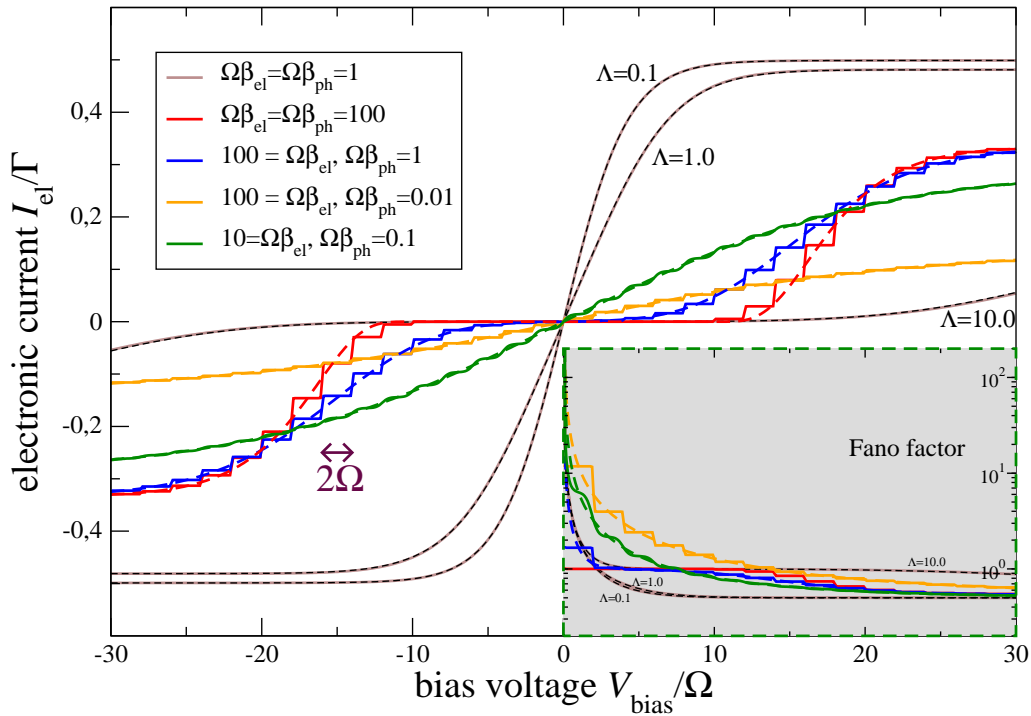}
\end{center}
\caption{\label{Fcurfanoel}
Electronic current and Fano factor (inset) for the single-mode (solid) and continuum (dashed) cases for different temperature configurations (colored curves) and different coupling strengths (brown solid and black dashed).
Low electronic temperatures are required to resolve the single-mode nature of the phonon bath (steplike solid curves vs. smooth dashed curves).
The reduction of even the large-bias currents is a consequence of a constant and finite width of the electronic tunneling rates.
Other Parameters: $\Gamma_L = \Gamma_R=\Gamma$, $\delta=10 \Omega$, $\epsilon/\Omega=0$, $\Omega=\omega_{\rm c}$, and for the bold colored curves $\Lambda=\abs{h}^2/\Omega^2=J_0=5$.
}
\end{figure}

The cumulants of the FCS such as electronic current 
$I_{\rm el} = \frac{d}{dt} \lim\limits_{t\to\infty} \expval{n_{\rm el}^{(R)}}$ and noise 
$S_{\rm el} = \frac{d}{dt} \lim\limits_{t\to\infty} \left(\expval{{n_{\rm el}^{(R)}}^2}-\expval{n_{\rm el}^{(R)}}^2\right)$
can be easily calculated using Eqs.~(\ref{Elmatfull}) and~(\ref{Eredlmat}) -- see e.g. the appendix.
We plot current and Fano factor $F=S/\abs{I}$ in Fig.~\ref{Fcurfanoel}.
At low electronic temperatures, and in the single-mode case, the phonon frequency can be deduced from the steps in the electronic current.
Generally, for strong SET-phonon coupling one observes a suppression of the finite-bias current in comparison to the uncoupled case -- 
known as Franck-Condon blockade~\cite{koch2005a}, which normally requires to treat phonons as dynamical degrees of freedom.
Remarkably, our simple $2\times 2$ rate equations even recover that the giant Fano factors typical for Franck-Condon blockade can only be observed when
electron and phonon temperatures are chosen very different, i.e., when the reservoirs are unequilibrated.
This behavior is found in the continuum phonon case (cf. dashed curves in Fig.~\ref{Fcurfanoel}) as well as in the 
large bias regime ($f_L(\omega)\to 1$ and $f_R(\omega)\to 0$ but keeping the energy-dependent tunneling rates).
In both wide-band and infinite bias regime we simply recover the conventional current of the SET (not shown). 

\begin{figure}[ht]
\begin{center}
\includegraphics[width=0.75\textwidth,clip=true]{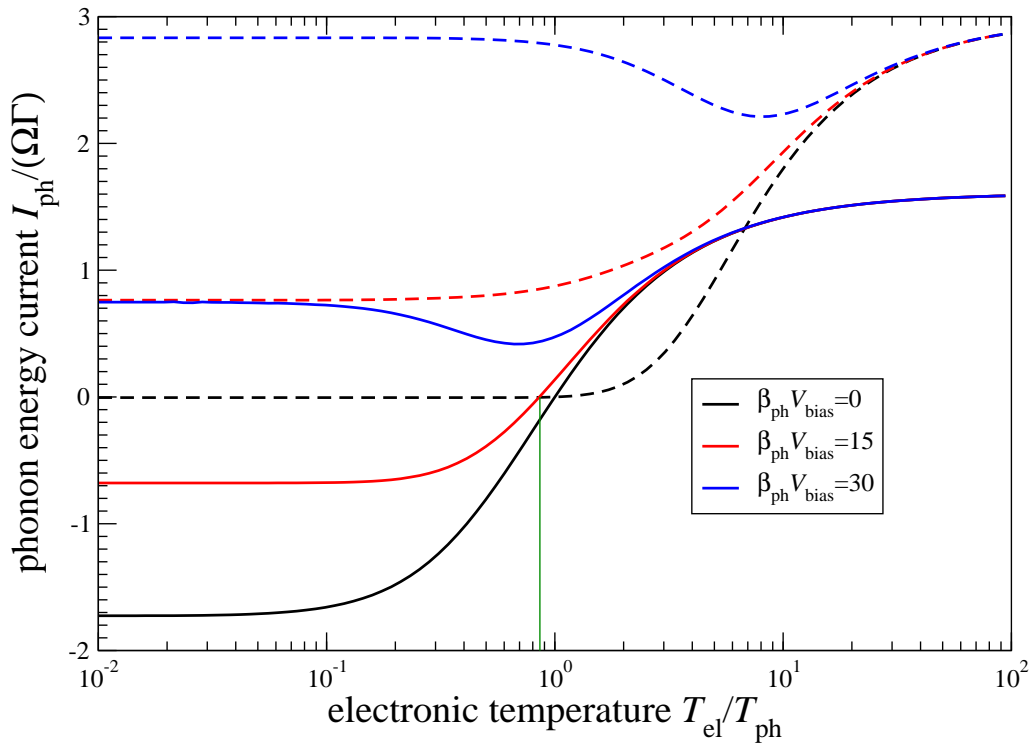}
\end{center}
\caption{\label{Fcurfanoph}
Total energy current into the phonon bath versus electronic temperature $T_{\rm el}/T_{\rm ph}$
for different phonon temperatures $\beta_{\rm ph}\Omega=0.1$ (solid), $\beta_{\rm ph}\Omega=1.0$ (dashed) and vanishing $V_{\rm bias}\Omega=0$ (black), intermediate
$V_{\rm bias}\Omega=15$ (red), and high $V_{\rm bias}\Omega=30$ (blue) electronic bias voltages.
Other Parameters: $\Gamma_L = \Gamma_R$, $\delta=10\Omega$, $\epsilon/\Omega=0$, $\Omega\beta_{\rm ph}=1$, $\abs{h}^2/\Omega^2=5$.
}
\end{figure}

A common problem in molecular transport spectroscopy is that the electronic current used to probe the phonon frequency will, at high bias, induce 
vibronic excitations, which may eventually lead to the destruction of the molecule~\cite{simine2012a}.
The low-dimensional model proposed here is quite useful to find regimes allowing electronic spectroscopy (finite bias) simultaneously with a net 
negative phonon energy current (implying a lowering of the effective molecular temperature), 
see the red solid curve in Fig.~\ref{Fcurfanoph}.
There, we evaluate the statistics of total phonon heat emission in the single-mode case from Liouvillian~(\ref{Elmatfull}) 
by not making any difference between right and left associated electronic jumps, see the appendix.
The particular parametrization in Fig.~\ref{Fcurfanoph} demonstrates a stationary configuration with only slightly different electronic and phonon temperatures 
$T_{\rm el}\approx 0.86 T_{\rm ph}$ at finite bias
(intersection point of the solid red curve with $I=0$ marked by the green vertical line), 
at which the phonon energy may be probed without inducing further heating.
More generally, determining the phonon frequency by electronic spectroscopy in a non-destructive manner 
requires electron and phonon temperatures where, within a finite voltage range, the phonon heat
current is non-positive.
A major advantage of our model is that this range can be estimated without 
having to consider hundreds or thousands of phonon states.


\section{Summary}

We studied the combined Full Counting Statistics of electrons and phonons for 
a special case of the Anderson-Holstein model using an extremely efficient 
two-dimensional rate equation for the electronic populations which treats 
the phonon mode as a strongly-coupled reservoir. 
Despite its simplicity, our approach was able to predict strong 
signatures of Franck-Condon blockade in the electronic current and noise.
Furthermore, we proved its thermodynamic consistency by explicitly deriving 
a universal fluctuation theorem for the entropy production and identifying
the three thermodynamic affinities characterizing our setup and imposing
universal symmetries on the Full Counting Statistics.
Our results hold for arbitrary electron-phonon coupling and also for 
multiple phonon modes as long as they are held at thermal equilibrium.
In addition, we found a non-universal fluctuation theorem referring only 
to the electronic counting statistics, which may be used to probe the 
phonon temperature.
The universal fluctuation theorem is recovered in this case when all 
terminal temperatures are set equal, which implies that the energy 
flow between the system and the phonons vanishes.
Due to its small dimensionality, our model may prove useful to identify 
appropriate regimes for performing non-destructive molecular spectroscopy.
Furthermore, we expect our approach to be also applicable to electronic 
transport through more complex structures.


\section{Acknowledgments}

Financial support by the DFG (SCHA 1646/2-1, SFB 910, and GRK 1558) and the National 
Research Fund, Luxembourg (project FNR/A11/02) is gratefully acknowledged.
The authors have profited from discussions with G. Kiesslich and P. Strasberg.
\\



\appendix

\section{Bath correlation functions for discrete phonon modes}\label{SAbcf}

Obviously, the phonon correlation function Eq.~(\ref{Ebcfphonon}) can be written as a product of single-mode correlation functions
\bea
C_{\rm ph}(\tau) = \prod_{q=1}^{Q} C_{\rm ph}^q(\tau)\,,
\eea
where $Q$ denotes the number of different phonon modes.
Each factor can be formally expanded in the variables $e^{\pm \ii\omega_q\tau}$ 
\bea
C_{\rm ph}^q(\tau) &=& e^{-\frac{\abs{h_q}^2}{\omega_q^2}\left(1+2\NB^q\right)} 
\sum_{m,m'=0}^\infty \left(\frac{\abs{h_q}^2}{\omega_q^2}\right)^{m+m'}
\frac{\left(\NB^q\right)^m \left(1+\NB^q\right)^{m'}}{m! m'!} e^{+\ii(m-m')\omega_q\tau}\nn
&\equiv& \sum_{m,m'=0}^\infty C_{mm',\rm ph}^q(\tau)\,.
\eea
Now, the key observation is that the terms in the sum can be interpreted as accounting 
for the emission (absorption) by the phonons of $m$ ($m'$) quanta with frequency $\omega_q$. 
Since we are interested in the net number 
of quanta absorption by the phonon bath $n=m'-m$, it is natural to define
\bea
C_{n,\rm ph}^q(\tau) \equiv \sum_{m,m'}^\infty \delta(m'-m,n) C_{mm',\rm ph}^q(\tau)
\eea
with $\delta(m,m')$ denoting the Kronecker symbol.
The identity 
\bea
f(a,b,n) &=& 
\sum_{m,m'=0}^\infty \frac{a^{m} b^{m'}}{m! m'!} \delta(m'-m,n)\nn
&=&
\left\{\begin{array}{ccc}
\sum \limits_{m=0}^\infty \frac{a^{m}}{m!} \frac{b^{m+n}}{(m+n)!} & : & n \ge 0\\
\sum\limits_{m=-n}^\infty  \frac{a^{m}}{m!} \frac{b^{m+n}}{(m+n)!} & : & n < 0
\end{array}\right.\nn
&=& \left(\frac{b}{a}\right)^{n/2} {\cal J}_{n}\left(2 \sqrt{a b}\right)\,,
\eea
where ${\cal J}_n(x)$ denotes the modified Bessel function of the first kind, 
takes the summation boundaries into account properly and implies
\bea
C_{\rm ph}^q(\tau) &=& \sum_{n=-\infty}^{+\infty} C_{n,\rm ph}^q(\tau)\nn
&=& \sum_{n=-\infty}^{+\infty} e^{-\ii n \omega_q \tau}
e^{-\frac{\abs{h_q}^2}{\omega_q^2}\left(1+2\NB^q\right)} 
\left(\frac{1+\NB^q}{\NB^q}\right)^\frac{n}{2}\times\nn
&&\times {\cal J}_n\left(2 \frac{\abs{h_q}^2}{\omega_q^2} \sqrt{\NB^q(1+\NB^q)}\right)\,.
\eea
For the full phonon correlation function, we can separate the $\tau$-dependence as
\bea\label{Edefweight}
C_{\rm ph}(\tau) &=& \sum_{\f{n}} e^{-\ii \f{n}\cdot\f{\Omega} \tau} \prod_{q=1}^Q \Big[e^{-\frac{\abs{h_q}^2}{\omega_q^2}\left(1+2\NB^q\right)}\times\nn
&&\times \left(\frac{1+\NB^q}{\NB^q}\right)^\frac{n_q}{2} {\cal J}_{n_q}\left(2 \frac{\abs{h_q}^2}{\omega_q^2} \sqrt{\NB^q(1+\NB^q)}\right)\Big]\nn
&\equiv& \sum_{\f{n}} e^{-\ii \f{n}\cdot\f{\Omega} \tau} C_{\rm ph}^{\f{n}}\,,
\eea
where $\sum\limits_{\f{n}} \equiv \sum\limits_{n_1=-\infty}^{+\infty} \ldots \sum\limits_{n_Q=-\infty}^{+\infty}$ 
such that each phonon mode has a different summation index 
and $\f{n} \equiv (n_1,\ldots, n_Q)$, $\f{\Omega} \equiv (\omega_1,\ldots, \omega_Q)$ with $\f{n} \cdot \f{\Omega} = \sum_{q=1}^Q n_q \omega_q$.
This enables one to express the Fourier transform of the total (electron and phonon) bath correlation function
\bea
\gamma_{\ell}^\alpha(\omega) &=& \int d\tau C_{\ell}^{\alpha,\rm el}(\tau) C_{\rm ph}(\tau) e^{+\ii\omega\tau}\nn
&=& \sum_{\f{n}_\alpha} \int d\tau \left[\frac{1}{2\pi} \int d\omega' \gamma_{\ell}^{\alpha,\rm el}(\omega') e^{-\ii\omega'\tau}\right]
 e^{+\ii \left(\omega-\f{n}_\alpha\cdot\f{\Omega}\right) \tau} C_{\rm ph}^{\f{n}_\alpha}\nn
&=& \sum_{\f{n}_\alpha} \gamma_{\ell}^{\alpha,\rm el}\left(\omega-\f{n}_\alpha\cdot\f{\Omega}\right)  C_{\rm ph}^{\f{n}_\alpha}
\equiv \sum_{\f{n}_\alpha} \gamma_{\ell,\f{n}_\alpha}^\alpha(\omega)
\eea
as a weighted superposition of purely electronic Fourier transforms that are evaluated at phonon-shifted frequencies.
Each term $\gamma_{\ell,\f{n}_\alpha}^\alpha$ in the sum accounts for a single electronic jump across junction $\alpha$ triggering the net 
absorption by the phonons of $(n_{\alpha,1},\ldots,n_{\alpha,Q})$ quanta carrying a net energy of $e_{\alpha} = \f{n}_{\alpha} \cdot \f{\Omega}$.
In particular, for a single phonon mode ($Q=1$) this leads to Eq.~(\ref{Ecfunc}) in the manuscript.

The discussion in the manuscript however also applies for multiple phonon modes at the same temperature:
The bosonic symmetry relation $\NB^q = e^{-\beta_{\rm ph} \omega_q} \left(1+\NB^q\right)$ for example implies for the weight factors in Eq.~(\ref{Edefweight})
\bea
C_{\rm ph}^{-\f{n}_\alpha} =  e^{-\beta_{\rm ph} \f{n}_\alpha \cdot \f{\Omega}} C_{\rm ph}^{+\f{n}_\alpha}\,,
\eea
which can be used to show that the single-mode KMS relation~(\ref{Ekms}) in the manuscript straightforwardly generalizes to
\bea\label{Ekmsapp}
\gamma_{12,+\f{n}_\alpha}^\alpha(-\omega) &=& e^{-\beta_\alpha(\omega-\mu_\alpha+\f{n}_\alpha \cdot\f{\Omega})} e^{\beta_{\rm ph} \f{n}_\alpha \cdot\f{\Omega}}
\gamma_{21,-\f{n}_\alpha}^\alpha(+\omega)\,,
\eea
where $e_\alpha = \f{n}_\alpha \cdot \f{\Omega}$ is the total net energy absorbed by the phonon bath when an electronic transition with lead $\alpha$ occurs.
For the total entropy production, this total transferred energy $e_\alpha$ is relevant.
Therefore, when inserting counting fields in the Liouvillian it suffices to count the net bosonic energy transfer
triggered by electronic jumps across each junction instead of counting each phonon mode separately, such that Eq.~(\ref{Elmatfull})
in the manuscript generalizes to
\bea\label{Elmatfullapp}
{\cal L}(\chi,\xi_L,\xi_R) &=& \sum_{\alpha\in\{L,R\}}\sum_{\f{n}_\alpha}\\
&&\left(\begin{array}{cc}
-\gamma_{12,\f{n}_\alpha}^\alpha(-\tilde{\epsilon}) & \gamma_{21,\f{n}_\alpha}^\alpha(+\tilde{\epsilon})e^{+\ii\chi_\alpha}e^{+\ii e_\alpha \xi_\alpha}\\
+\gamma_{12,\f{n}_\alpha}^\alpha(-\tilde{\epsilon}) e^{-\ii\chi_\alpha}e^{+\ii e_\alpha \xi_\alpha} & -\gamma_{21,\f{n}_\alpha}^\alpha(+\tilde{\epsilon})
\end{array}\right)\nonumber
\eea
with the renormalized dot level $\tilde{\epsilon}=\epsilon-\sum\limits_{q=1}^Q \frac{\abs{h_q}^2}{\omega_q}$.
The dimensionless electronic counting fields $\chi_\alpha$ enable one to extract the complete electronic particle counting statistics.
In contrast, the phonon energy counting fields $\xi_\alpha$ have dimension of inverse energy and account for the statistics of phonon-related
energy transfers triggered by electronic jumps across junction $\alpha$.
For the case of a single phonon mode (see main manuscript), the total phonon energy and phonon number for electronic jumps across junction $\alpha$ are tightly coupled 
$e_\alpha = n_\alpha \Omega$, such that the full energy counting statistics for the phonons also yields the full particle counting statistics.
To obtain the full particle counting statistics for multi-mode phonons one would have to make the counting fields 
mode-dependent $e^{\ii \f{n}_\alpha\cdot\f{\Omega}\xi_\alpha} \to e^{\ii \sum_{q=1}^Q n_{\alpha,q} \xi_{\alpha,q}}$, which
would exceed the focus of this paper.

\section{Full Counting Statistics and the Fluctuation Theorem}\label{SAfcsft}

The counting-field dependent Liouvillian~(\ref{Elmatfullapp}) and its single-mode version~(\ref{Elmatfull}) enable for the construction of the FCS, i.e., the 
probability $P_{n_{\rm el}^{(L)},n_{\rm el}^{(R)},e_{\rm ph}^{(L)}, e_{\rm ph}^{(R)}}(t)$ to observe $n_{\rm el}^{(\alpha)}$ net 
electrons transfers to lead $\alpha$ and a net associated energy $e_{\rm ph}^{(\alpha)}$ absorbed by the phonon reservoir.
We are interested in the large-time limit, due to charge conservation, it suffices to consider a single electronic counting field.
In the following, we will therefore consider $\chi_L=0$ and $\chi_R=\chi$ leading to the probability distribution 
$P_{n_{\rm el}^{(R)},e_{\rm ph}^{(L)}, e_{\rm ph}^{(R)}}(t)$ associated with $n_{\rm el}^{(R)}$ net electrons emitted into the right reservoir.
It is technically convenient to construct cumulants of the probability distribution via the cumulant-generating function (CGF),
which in the long-term limit becomes
\bea\label{Elongtermcgf}
{\cal C}(\chi,\xi_L,\xi_R,t) \stackrel{t\to\infty}{\to} \lambda(\chi,\xi_L,\xi_R) t\,,
\eea
where $\lambda(\chi,\xi_L,\xi_R)$ denotes the dominant eigenvalue of the Liouvillian 
which vanishes when all counting fields are set to zero $\lambda(0,0,0)=0$.
Cumulants are obtained by taking derivatives with respect to the counting field of interest:
The trivial long-term time-dependence then enables one to consider the CGF for the currents $\lambda(\chi,\xi_L,\xi_R)$ instead,
such that the electronic particle current and noise at the right junction for example may be obtained via
\bea
I = (-\ii) \partial_\chi \left.\lambda(\chi,0,0)\right|_{\chi=0}\,,\qquad
S = (-\ii)^2 \partial_\chi^2 \left.\lambda(\chi,0,0)\right|_{\chi=0}\,.
\eea
Similarly, the total phonon energy current and noise follow by disregarding the difference between left- and right-associated emissions
\bea
I_{\rm ph} = (-\ii) \partial_\xi \left.\lambda(0,\xi,\xi)\right|_{\xi=0}\,,\qquad
S_{\rm ph} = (-\ii)^2 \partial_\xi^2 \left.\lambda(0,\xi,\xi)\right|_{\xi=0}\,.
\eea

The full distribution may be obtained from the CGF by performing an inverse Fourier transform, which in the long-term limit becomes
\bea
\hspace{-0.8cm}
P_{n_{\rm el}^{(R)},e_{\rm ph}^{(L)},e_{\rm ph}^{(R)}}(t) &\stackrel{t\to\infty}{\to}&
\int\limits_{-\pi}^{+\pi} \frac{d\chi}{2\pi}
\int\limits_{-\infty}^{+\infty} \frac{d\xi_L}{2\pi}
\int\limits_{-\infty}^{+\infty} \frac{d\xi_R}{2\pi}
e^{\lambda(\chi,\xi_L,\xi_R) t} e^{- \ii \left(n_{\rm el}^{(R)} \chi+e_{\rm ph}^{(L)} \xi_L+e_{\rm ph}^{(R)} \xi_R\right)}\,.
\eea
From properties of the Fourier transform it follows that a shift-symmetry of the CGF is associated with a fluctuation theorem
for the FCS, i.e., 
\bea
\lambda(-\chi,-\xi_L,-\xi_R)&=&\lambda(+\chi+\ii A,+\xi_L+\ii A_L,+\xi_R+\ii A_R)\nn
&\Longleftrightarrow&\nn
\lim\limits_{t\to\infty} \ln \frac{P_{+n_{\rm el}^{(R)},+e_{\rm ph}^{(L)},+e_{\rm ph}^{(R)}}(t)}{P_{-n_{\rm el}^{(R)},-e_{\rm ph}^{(L)},-e_{\rm ph}^{(R)}}(t)} 
&=& n_{\rm el}^{(R)} A + e_{\rm ph}^{(L)} A_L + e_{\rm ph}^{(R)} A_R\,.
\eea
We aim at obtaining the affinities $A$, $A_L$, and $A_R$, which with Eq.~(\ref{Elongtermcgf}) yield a long-term symmetry of the CGF.
In our model~(\ref{Elmatfullapp}), it is most convenient to derive the corresponding symmetry from the characteristic polynomial of the Liouvillian.

To see it we first write the off-diagonal matrix elements of the Liouvillian as explicit functions of the counting fields
(the diagonal entries do not depend on the counting fields)
\bea
{\cal L}_{12}(+\chi,+\xi_L,+\xi_R) &=& \sum_n\left(\gamma_{21,n}^L(+\tilde{\epsilon}) e^{+\ii n \Omega \xi_L}
+ \gamma_{21,n}^R(+\tilde{\epsilon}) e^{+\ii\chi} e^{+\ii n \Omega \xi_R}\right)\,,\nn
{\cal L}_{21}(+\chi,+\xi_L,+\xi_R) &=& \sum_n\left(\gamma_{12,n}^L(-\tilde{\epsilon}) e^{+\ii n \Omega \xi_L}
+ \gamma_{12,n}^R(-\tilde{\epsilon}) e^{-\ii\chi} e^{+\ii n \Omega \xi_R}\right)\,,
\eea
where we recall that $\chi_L=0$ and $\chi_R=\chi$.
Inserting the KMS condition~(\ref{Ekmsapp}) and replacing $n\to-n$ under the sum we obtain
\bea\label{Esym1app}
{\cal L}_{12}(-\chi,-\xi_L,-\xi_R) &=& \sum_n\Big(
\gamma_{12,n}^L(-\tilde{\epsilon}) e^{+\ii n \Omega \xi_L} e^{-(\beta_{\rm ph}-\beta_L) n \Omega} 
e^{\beta_L(\tilde{\epsilon}-\mu_L)}\nn
&&+\gamma_{12,n}^R(-\tilde{\epsilon}) e^{-\ii \chi} e^{+\ii n \Omega \xi_R} e^{-(\beta_{\rm ph}-\beta_R) n \Omega} 
e^{\beta_R(\tilde{\epsilon}-\mu_R)}\Big)\nn
&=& e^{+\beta_L(\tilde{\epsilon}-\mu_L)} 
{\cal L}_{21}(\chi+\ii A, \xi_L+\ii A_L, \xi_R+\ii A_R)
\eea
with $A=\beta_R(\tilde{\epsilon}-\mu_R)-\beta_L(\tilde{\epsilon}-\mu_L)$, 
$A_L = \beta_{\rm ph}-\beta_L$,  and
$A_R = \beta_{\rm ph}-\beta_R$.
This symmetry is equivalent to
\bea\label{Esym2app}
{\cal L}_{21}(-\chi,-\xi_L,-\xi_R) &=& e^{-\beta_L(\tilde{\epsilon}-\mu_L)} 
{\cal L}_{12}(\chi+\ii A, \xi_L+\ii A_L, \xi_R+\ii A_R)\,.
\eea
The characteristic polynomial can be written as 
\bea
{\cal D}(\chi,\xi_L,\xi_R) &=& \abs{{\cal L}(\chi,\xi_L,\xi_R) - \lambda \f{1}}\nn
&=& \left({\cal L}_{11}-\lambda\right)\left({\cal L}_{22}-\lambda\right) 
- {\cal L}_{21}(\chi,\xi_L,\xi_R) {\cal L}_{12}(\chi,\xi_L,\xi_R)\,,
\eea
where we see that when evaluating the expression at negative arguments the exponential prefactors
from Eqns.~(\ref{Esym1app}) and~(\ref{Esym2app}) cancel and we easily read off the symmetry
\bea
{\cal D}(-\chi,-\xi_L,-\xi_R) &=& {\cal D}(+\chi+\ii A,+\xi_L+\ii A_L,+\xi_R+\ii A_R)\,.
\eea
Since the two eigenvalues of the Liouvillian $\lambda(\chi,\xi_L,\xi_R)$ and $\bar{\lambda}(\chi,\xi_L,\xi_R)$ are given by the roots of the characteristic polynomial
${\cal D}(\chi,\xi_L,\xi_R) = \left[\lambda-\lambda(\chi,\xi_L,\xi_R)\right]\left[\lambda-\bar{\lambda}(\chi,\xi_L,\xi_R)\right]$, this symmetry transfers 
to the long-term CGF as
\bea
\lambda(-\chi,-\xi_L,-\xi_R) &=& \lambda(+\chi+\ii A,+\xi_L+\ii A_L, +\xi_R+\ii A_R)
\eea
and eventually implies the validity of the full fluctuation theorem~(\ref{Efttotal}) in the main text.

\end{document}